\documentclass[aps,superscriptaddress,groupedaddress,twocolumn,floatfix,10pt]{revtex4}
\usepackage{natbib}
\usepackage{graphicx} % needed for figures
\usepackage{dcolumn} % needed for some tables
\usepackage{bm} % for math
\usepackage{amssymb,amsmath} % for math
\usepackage{textcomp} % for text
\usepackage[utf8x]{inputenc}
\usepackage{braket}
\usepackage{amsmath}
\usepackage{placeins}
\usepackage{xcolor}
\usepackage{pdfpages}
\usepackage{soul}
\usepackage{hyperref}       % hyperlinks
\usepackage{url}            % simple URL typesetting
\usepackage{booktabs}       % professional-quality tables
\usepackage{todonotes} %\todo[prepend/inline]{ }
\usepackage{changebar}
\usepackage{changes}
\definechangesauthor{AB}
\begin{document}

\pagestyle{empty}

\title{Stock price formation: useful insights from a multi-agent reinforcement learning model}

\author{
Johann Lussange\\
Laboratoire des Neurosciences Cognitives, INSERM U960, Département des Études Cognitives\\
École Normale Supérieure, 29 rue d'Ulm, 75005, Paris, France.\\
Sacha Bourgeois-Gironde\\
Institut Jean-Nicod, UMR 8129, Département des Études Cognitives,\\
École Normale Supérieure, 29 rue d'Ulm, 75005, Paris, France.\\ 
Laboratoire d'Économie Mathématique et de Microéconomie Appliquée, EA 4442\\
Université Paris II Panthéon-Assas, 4 rue Blaise Desgoffe, 75006, Paris, France.\\
Stefano Palminteri\\
Laboratoire des Neurosciences Cognitives, INSERM U960, Département des Études Cognitives\\
École Normale Supérieure, 29 rue d'Ulm, 75005, Paris, France.\\
Boris Gutkin\\
Laboratoire des Neurosciences Cognitives, INSERM U960, Département des Études Cognitives\\
École Normale Supérieure, 29 rue d'Ulm, 75005, Paris, France.\\ 
Center for Cognition and Decision Making, Department of Psychology\\
NU University Higher School of Economics, 8 Myasnitskaya st., 101000, Moscow, Russia.\\
}

%\date{\today}

\begin{abstract} 
In the past, financial stock markets have been studied with previous generations of multi-agent systems (MAS) that relied on zero-intelligence agents, and often the necessity to implement so-called noise traders to sub-optimally emulate price formation processes. However recent advances in the fields of neuroscience and machine learning have overall brought the possibility for new tools to the bottom-up statistical inference of complex systems. Most importantly, such tools allows for studying new fields, such as agent learning, which in finance is central to information and stock price estimation. We present here the results of a new generation MAS stock market simulator, where each agent autonomously learns to do price forecasting and stock trading via model-free reinforcement learning, and where the collective behaviour of all agents decisions to trade feed a centralised double-auction limit order book, emulating price and volume microstructures. We study here what such agents learn in detail, and how heterogenous are the policies they develop over time. We also show how the agents learning rates, and their propensity to be chartist or fundamentalist impacts the overall market stability and agent individual performance. We conclude with a study on the impact of agent information via random trading. 
\end{abstract}

\maketitle

%PhySH concepts: Stochastic processes, Dynamics of networks, Network phase transitions, Self-organized systems, Network formation & growth, Nonequilibrium statistical mechanics

%- Policies (F, T)
%- Learning rates (percent, scalar)
%- Strategies (reflexivity, gesture)
%- Mesoscales (best, worst, noise)

\section{Introduction}

\textit{Background}: Multi-agent systems (MAS) or agent-based models (ABM) have had a long history of statistical inference in quantitative finance research. In particular, they have been used to study phenomena leading to price formation and hence general market microstructure, such as: the law of supply and demand~\cite{Benzaquen2018}, game theory~\cite{ErevRoth2014}, order books~\cite{Huang2015}, high-frequency trading~\cite{Wah2013,Aloud2014}, cross-market structure~\cite{Xu2014}, quantitative easing~\cite{Westerhoff2008}, market regulatory impact~\cite{Boero2015}, or other exogenous effects~\cite{Gualdi2015}. Remarkably, financial MAS have highlighted over the years certain specific patterns that are proper to virtually almost all asset classes and time scales, called \textit{stylised facts}. These have since been an active topic of quantitative research~\cite{Lipski2013,Barde2015}, and can be grouped in three main neighbouring categories: i- distributions of price returns are non-gaussian~\cite{Cristelli2014,Cont2001,Potters2001} (they are asymmetric, negatively skewed, and platykurtic), ii- volatilities and volumes are clustered~\cite{Lipski2013} (large jumps in prices and volumes are more likely to be followed by the same), iii- price auto-correlations decay~\cite{Cont2001,Cont2005} (there is a loss of arbitrage opportunity). In particular, the second stylised fact has long-range implications on the dynamics of meta-orders, and comprises the \textit{square-root impact law}~\cite{Bouchaud2018} (growth in square-root of orders impact with traded volumes). Implicit consequences of these stylised facts have fed numerous discussions pertaining to the validity of market memory~\cite{Cont2005,Cristelli2014} and the extension of the efficient market hypothesis~\cite{Fama1970,Bera2015}. 

\vspace{1mm}

\textit{Epistemology}: In the past, the epistemological pertinence of such models was sometimes put into question, because of the general conceptual challenge of framing realistic agents. Unlike other disciplines from hard science, financial MAS indeed had to justify their proper bottom-up approach to complex system inference~\cite{Gao2011} in the light of a specific challenge of their field, namely the difficulty to realistically model and emulate human agents. Such critics especially carried greater weight by the fact that previous generations of financial MAS relied on so-called zero-intelligence agents~\citep{Gode1993}: these would trade according to specific and imbedded rules of trading, thereby overshadowing certain dynamics proper to game and decision theory that are crucial to real market activity. Yet, if compared with other famous types of models used in quantitative finance, like econometrics~\cite{Greene2017}, MAS have two major advantages: i- they naturally display specific emergent phenomena proper to complex systems~\cite{Bouchaud2019}, and ii- they require fewer model assumptions (no gaussian distributions, no efficient market hypothesis~\cite{Fama1970,Bera2015}, etc.). As for their shortcomings, we can mention specific conceptual challenges pertaining to: i- modelling complex system heterogeneity~\cite{Chen2017}, ii- discretionary framing of certain model parameters (e.g. number of agents) apart from possible empiricism~\cite{Platt2018}. 

\vspace{1mm}

\textit{Prospects}: These general conceptual and epistemological considerations have been upset in the past few years, by the notable progress of two fields. The first, namely machine learning and especially multi-agent reinforcement learning~\cite{Silver2018} (RL), has produced spectacular results that have far-reaching applications to other domains of interest to quantitative finance, like decision theory and game theory. The second, neuroscience and neurofinance especially, has benefitted from the wide use of brain-imaging devices~\cite{Eickhoff2018} and peer data curation~\cite{Smith2018}. On both ends, some sort of technological emergence within these two fields is of special interest and relevance to financial MAS research, in that machine learning can imbed results from the latter~\cite{Lefebvre2017,Palminteri2015}, and vice-versa~\cite{Duncan2018,Momennejad2017}. Among other machine learning applications to finance~\citep{Hu2019,Neuneier1997,Deng2017}, and in a way similar to what has been done for recent order book models~\citep{Spooner2018,Biondo2019,Sirignano2019}, next-generation MAS stock market simulators can now be designed, and outstrip the former epistemological considerations pertaining to agent design, with a whole new degree of realism in market microstructure emulation. More importantly, these can address issues never computationally studied before, such as the crucial issues of agent information and learning, central to price formation~\cite{Dodonova2018,Naik2018} and hence to all market activity. 

\vspace{1mm}

\textit{Our study}: We have designed such a MAS stock market simulator, in which agents autonomously learn to do price forecasting and stock trading by reinforcement learning, via a centralised double-auction limit order book, and shown in a previous work~\cite{Lussange2019} its calibration performances with respect to real market data from the London Stock Exchange daily quotes between $2008$ and $2018$. We shall not review here its design in full details, but will recall its general architecture in Section II. Then our study will focus in Section III on agent learning: how can we gauge how heterogenous are the agent policies as compared to one another, and what are the trading characteristics of successful agents. Finally in Section IV, we will study the impact of such agent learning on the whole market at the mesoscale, and in particular with respect to herding or reflexivity effects when agents emulate the investments of the best (or worst) agent, and when increasing proportions of agents trade randomly as ``noise traders." As said, the greatest interest of financial MAS is to study and quantitatively gauge the impact of agent learning, which is at the heart of the price formation processes and hence of all market activity, and the motivation for multi-agent reinforcement learning for such a framework is motivated by the important parallels between reinforcement learning and decision processes in the brain~\cite{Dayan2008}.

% This section comes the PRE article!
\section{Model}

\textit{Architecture}: We here briefly sketch the general architecture of our MAS simulator. Our model relies on a number $I$ of economic agents autonomously trading a number $J$ of stocks, over a number $T$ of simulation time steps, where we set $T_{y}=286, T_{m}=21, T_{w}=5$ as the number of annual trading days. All $I$ agents and their individual parameters are initialised at $t=0$, with a portfolio made of specific stock holdings of value $A_{equity}^{i}(t)=\sum_{j=0}^J Q^{i,j}(t)P^{j}(t)$, where $Q^{i,j}(t)$ is the number of stocks $j$ of agent $i$ and $P^{j}(t)$ the market price of stock $j$, together with risk-free assets (e.g. a bank account) of value $A_{bonds}^{i}(t)$. Then all market prices are initialised at $P^{j}(t=0)=\pounds 100$, and as in other models~\cite{Franke2011,Chiarella2007}, the simulation generates $J$ time series $\mathcal{T}^{j}(t)$, which correspond to the fundamental values of the stocks. These are not fully known by the $I$ agents. Instead, each agent $i$ approximates the values $T^j(t)$ of stock $j$ according to a proprietary rule~\citep{Murray1994} of cointegration $\kappa ^{i,j} [ \mathcal{T}^{j}(t) ]=\mathcal{B}^{i,j}(t)$. The time series $\mathcal{B}^{i,j}(t)$ are hence the approximation of the fundamental values of stock $j$ over time $t$ according to agent $i$. Each agent thus relies on these two sources of information for its stock pricing strategy: one that is chartist, and one that is fundamental. At $t=0$, the agent are agnostic with respect to trading. They are then allowed to learn over the course of $1000$ time steps, after which their stock holdings and risk-free assets are reset back to their first value. The simulation and following results are hence applied to agents after this learning phase of $1000$ time steps. 

\textit{Initialisation}: Let $\mathcal{U} ()$ and $\mathcal{U} \{ \}$ denote the continuous and discrete uniform distributions, respectively. Each agent is then initialised with the following parameters: a drawdown limit $l^{i} \sim \mathcal{U} (50 \%, 60\%)$ (which is the maximum year-to-date loss in net asset value below which the agent is set as bankrupt), a reflexivity parameter $\rho^{i} \sim \mathcal{U} (0, 100\%)$ (which gauges how fundamental or chartist the agent is via a weighted average of its price forecast), an investment horizon $\tau^{i} \sim \mathcal{U} \{T_w, 6T_m \}$ (which is the number of time steps after which the agent liquidates its position), a trading window $w^{i} \sim \mathcal{U} \{T_w, \tau^{i} \}$ (which assesses the optimal trading time for sending an order), a memory interval $h^{i} \sim \mathcal{U} \{T_w, T-\tau^{i}-2T_w \}$ (which is the size of the past lag interval used by the agent for its learning process), a transaction gesture $g^{i} \sim \mathcal{U} (0.2, 0.8)$ (which scales with bid-ask spread to set how far above or below the value of its own stock pricing the agent is willing to deal the transaction), and a reinforcement learning rate $\alpha \sim \mathcal{U} (0.05, 0.20)$ proper to both reinforcement learning algorithms $\mathcal{F}^{i}$ and $\mathcal{T}^{i}$ (see below).

\textit{Order book}: At each time step, the agents may send transaction orders to the order book of each stock, whose function is to match these orders and process associated business transactions. More specifically, a number $J$ of order books are filled with all the agents' trading limit orders for each stock $j$ at time step $t$. All buy orders are there sorted by descending bid prices, all sell orders are sorted by ascending ask prices, each with their own associated number of stocks to trade. Then the order book clears these matching orders at same time step $t$, with each transaction set at mid-price between buy and sell-side, starting from the top of the order book to the lowest level where the bid price still exceeds the ask price. Importantly, we then define the market price $P^{j}(t+1)$ of stock $j$ at the next time step $t$ as that last and lowest level mid-price cleared by the order book. We also define the trading volume $V^{j}(t+1)$ as the number of stocks $j$ traded during that same time $t$. We also model the friction costs via broker fees~\citep{IG} applied to each transaction set at $0.1 \%$, an annual risk-free rate of $1 \%$ applied to $A_{bonds}^{i}(t)$, and an annual stock dividend yield of $2 \%$ according to~\citep{DividendYield} applied to $A_{equity}^{i}(t)$. 

\textit{Agents}: Each agent autonomously uses two distinct reinforcement learning algorithms to interact with the market. For a brief introductory sum up of reinforcement learning, we refer the reader to~\cite{Lussange2019}, and to~\cite{SuttonBarto,Wiering2012,Csaba2010} for a thorough study of the subject. A first algorithm $\mathcal{F}^{i}$ learns the optimal econometric prediction function for the agent's investment horizon, depending on specific local characteristics of the market microstructure and the agent's fundamental valuation $\mathcal{B}^{i,j}(t)$. It thus outputs this price forecast, which will in turn enter as input the second reinforcement learning algorithm $\mathcal{T}^{i}$. This second algorithm is in charge of sending an optimal limit order to a double auction order book~\citep{Mota2016} at this same time step, based on this prediction and a few other market microstructure and agent portfolio indicators. Each reinforcement learning algorithm is individually ran by each agent $i$ following a direct policy search, for each stock $j$, and at each time step $t$. Each algorithm has $27 \times 27=729$ and $108 \times 9=972$ potential action-state pairs, respectively. We define the sets of states $\mathcal{S}$, actions $\mathcal{A}$, and returns $\mathcal{R}$ of these two algorithms according to the following: 

i- \textit{Forecasting}: In the first algorithm $\mathcal{F}^{i}$, which is used for price forecasting, the agent continuously monitors the longer-term volatility of the stock prices $s_0^{\mathcal{F}}=\{0, 1, 2\}$, their shorter-term volatility $s_1^{\mathcal{F}}=\{0, 1, 2\}$, and the gap between its own present fundamental valuation and the present market price $s_2^{\mathcal{F}}=\{0, 1, 2\}$. This allows the agent to retrieve useful information on the microstructure and topology of the volatility, while avoiding dimensionality issues. Out of this state, it learns to optimise its price prediction at its investment horizon $\tau^{i}$ by selecting from a direct policy search three possible actions: choosing a simple forecasting econometric tool based on mean-reverting, averaging, or trend-following market prices $a_0^{\mathcal{F}}=\{0, 1, 2\}$, choosing the size of the historical lag interval for this forecast $a_1^{\mathcal{F}}=\{0, 1, 2\}$, and choosing the weight of its own fundamental stock pricing in an overall future price estimation, that is both fundamentalist and chartist $a_2^{\mathcal{F}}=\{0, 1, 2\}$. This is done in proportion to the agent reflexivity parameter $\rho^{i}$. Via this action $a_2^{\mathcal{F}}$, the agent thus learns to gauge how fundamental or chartist it should be is in its price valuation, via a weighted average involving $\rho^{i}$ in the agent's technical forecast of the market price $\hat{P}^{i,j}(t)$ and its fundamental pricing $\mathcal{B}^{i,j}(t)$. At each time step, the rewards $r^{\mathcal{F}}=\{-4, -2, -1, 1, 2, 4 \}$ are defined according to percentiles in the distribution of the agent's mismatches between past forecasts at time $t-\tau^{i}$ and their eventual price realisation at time $t$. In parallel, an off-policy method computes the optimal action that was to be performed $t-\tau^{i}$ time steps ago, now that the market price $P^{j}(t)$ is realised, and updates the agent policy accordingly. 

ii- \textit{Trading}: In the second algorithm $\mathcal{T}^{i}$, which is used for stock trading, the agent continuously monitors whether the stock prices are increasing or decreasing according to the output of the former algorithm $s_0^{\mathcal{T}}=\{0, 1, 2\}$, their volatility $s_1^{\mathcal{T}}=\{0, 1, 2\}$, its risk-free assets $s_2^{\mathcal{T}}=\{0, 1\}$, its quantity of stock holdings $s_3^{\mathcal{T}}=\{0, 1\}$, and the traded volumes of stock $j$ at former time step $s_4^{\mathcal{T}}=\{0, 1, 2\}$. Out of this state, it learns to optimise its investments by selecting two possible actions via a direct policy search: sending a transaction order to the order book as holding, buying, or selling a position in a given amount proportional to its risk-free assets and stock holdings $a_0^{\mathcal{T}}=\{0, 1, 2\}$, and at what price wrt. the law of supply and demand $a_1^{\mathcal{T}}=\{0, 1, 2\}$. The cashflow difference between the profit or loss consequent to the agent's action, and that without action having been taken, are then computed $\tau^{i}$ time steps after each transaction. The rewards $r^{\mathcal{T}}=\{-4, -2, -1, 1, 2, 4 \}$ are defined according to percentiles in the distribution of these agent's past cashflow differences. In parallel, an off-policy method computes the optimal action that was to be performed $t-\tau^{i}$ time steps ago, now that the market price $P^{j}(t)$ is realised, and updates the agent policy according to it.

\section{Agent learning}

\subsection{Agent policy heterogeneity}

We first want to measure the heterogeneity of the individual policies of these best and worse agents, not only as compared to one another as groups, but also to all other agents in the whole market population, and dynamically as a function of time. For this we compute matrices $\mathcal D^{\mathcal{F}}$ and $\mathcal D^{\mathcal{T}}$ of dimensions $I \times I$, where each matrix element $d^{\mathcal{F}}_{m, n}$ and $d^{\mathcal{T}}_{m, n}$ respectively, is computed as the average of the absolute differences between the probabilities $p_{i_m}$ and $p_{i_n}$ within the policy of agents $i_m$ and $i_n$, for each policy $\pi^{\mathcal{F}}(s,a)$ and $\pi^{\mathcal{T}}(s,a)$: 

\begin{eqnarray}
\label{Equ1}
d_{m, n}=\frac{1}{|\mathcal{S}| |\mathcal{A}|} \sum_{k_s=1}^{\mathcal{S}} \sum_{k_a=1}^{\mathcal{A}} | p_{i_m}(s_{k_s}, a_{k_a}) - p_{i_n}(s_{k_s}, a_{k_a}) |
\end{eqnarray}

\begin{figure}[!htbp]
\begin{centering}
\includegraphics[scale=0.53]{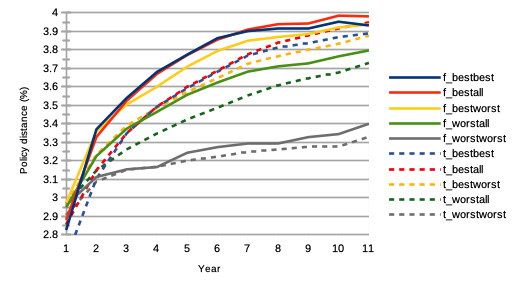}
\caption{\label{K1} Policy distances as defined by equation \ref{Equ1} for the first algorithm $\mathcal{F}()$ (continuous curves) and the second algorithm $\mathcal{T}()$ (dashed curves), scaled to $\mathcal{F}()$ values to account for different numbers of state-action pairs, as a function of simulation time in years, expressing in percentage the heterogeneity between different groups of agents : best $10\%$ agents with themselves (blue), best $10\%$ agents with all other agents (red), best $10\%$ agents with with worst $10\%$ agents (yellow), worst $10\%$ agents with all other agents (green), and worst $10\%$ agents with themselves (brown). The simulations are generated with parameters $I=500$, $J=1$, $T=2875$, $S=20$.}
\end{centering}
\end{figure}

The results are shown on Fig. \ref{K1}, where the values corresponding to $\mathcal{T}()$ were scaled to those corresponding to $\mathcal{F}()$, in order to account for the different numbers of state-action pairs. For the first forecasting algorithm $\mathcal{F}()$, we see that the best performing agents together converge to a pool of more diverse forecasting strategies, as compared to one another, but also to the worst performing agents and the rest of all market agents. Interestingly, the worst performing agents are less disparate in their policies among themselves, even more so than with regards to the rest of the market population. We see similar dynamics with the trading algorithm $\mathcal{T}()$, with an even more pronounced heterogeneity among groups. One can say that as simulation time passes, worst performing agents have more in common among themselves, than best performing agents among themselves. This is a remarkable prediction under our model assumptions, because of its implication to trading strategies: there are more ways to succeed than to fail. From a regulation point of view, this also potentially implies that financial stock markets benefit in stability from the multitudes of available trading instruments, structured products, and a diversification of investment strategies. Finally, we also note that the curves corresponding to the forecasting algorithm $\mathcal{F}()$ are sorted like those of the trading algorithm $\mathcal{T}()$, and nearly overlap one another. We posit this to be a consequence of the fact that each agent has identical reinforcement learning parameters (learning rate, rolling intervals, etc.) for both algorithms $\mathcal{F}()$ and $\mathcal{T}()$.

\subsection{Agent trading strategy}

\begin{figure}[!htbp]
\includegraphics[scale=0.53]{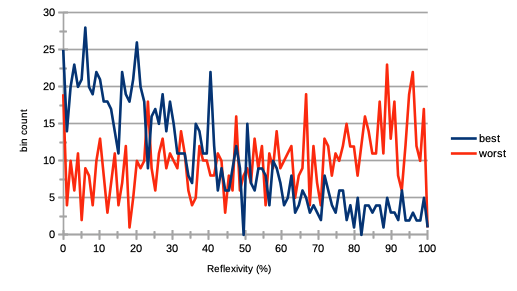}
\caption{\label{I1} Distribution of the reflexivity parameter $\rho^{i}$ of the $10 \%$ best (blue curve) and $10 \%$ worse (red curve) agents, at the end of the simulation set with parameters $I=500$, $J=1$, $T=2875$, $S=20$.}
\end{figure}

One of the first valuable statistical inference from the simulation is to gauge the agent propensity to engage in fundamentalist or chartist stock valuation for trading. At the mesoscale, this is one of the greatest known source of bubble formation and other so-called reflexivity effects~\cite{Hardiman2013}. Recall from Section II that each agent was initialised with a reflexivity parameter $\rho^{i} \sim \mathcal{U} (0, 100\%)$, reflecting how fundamental or chartist the agent is in its asset valuation. This reflexivity parameter $\rho^{i}$ works with a weighted average between the agent's technical price forecast and its cointegrated estimation of the fundamental value of the stock. The agent learns to optimise this parameter through action $a_2^{\mathcal{F}}$, and hence learns to be more chartist or fundamentalist, depending on the market dynamics, represented by its states $s_0^{\mathcal{F}}$ (longer-term volatility of the stock prices), $s_1^{\mathcal{F}}$ (shorter-term volatility), and $s_2^{\mathcal{F}}$ (gap between the agent's own present fundamental valuation and present market price). We show the results on Fig. \ref{I1}, where we can see at the end $t=T$ of the simulation a trend for the $10 \%$ best agents to be fundamentalists and the $10 \%$ worse agents to be chartists. We see that best performing agents have a tendency for being more fundamentalist, while the worst performing agents have for being more chartists. This can be linked with assumptions of the efficient market hypothesis of~\cite{Fama1970}, if we would consider that all the information that is endogenous to the market is retrieved by the agents (in the sense that it is available to all), but that exogenous information would be less accessible, and hence a filter to screen agent trading performance. 

\begin{figure}[!htbp]
\includegraphics[scale=0.53]{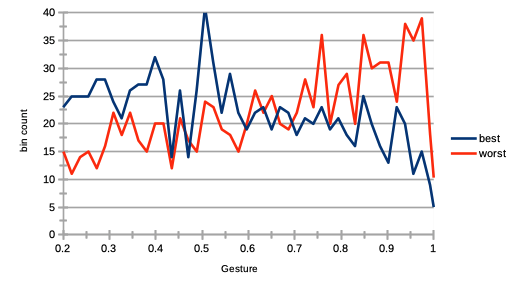}
\caption{\label{I2} Distribution of the gesture parameter $g^{i}$ of the $10 \%$ best (blue curve) and $10 \%$ worse (red curve) agents, at the end of the simulation set with parameters $I=500$, $J=1$, $T=2875$, $S=20$.}
\end{figure}

Another minor trait of reflexivity at the agent level is how much agents differ in their price estimation, and hence in the bid-ask spread formation. We can have a look into this spread and price formation process by checking the propensity for best performing agents to have a large gesture, as compared to worst performing ones. Such study has many parallels with market making and other strategies based on scalping the bid-ask spread. Recall each agent was initialised with a transaction gesture $g^{i} \sim \mathcal{U} (0.2, 0.8)$, reflecting how far above or below its own asset pricing the agent is willing to trade. The results are shown on Fig. \ref{I2}, where we can see that best performing agents have a propensity for having a smaller transaction gesture (propensity to bid larger prices and ask smaller prices in transaction orders), while worst performing agents have a propensity for having a larger one (propensity to bid smaller prices and ask larger prices in transaction orders). Interestingly, such results indicate that ``tougher" negotiators would thus statistically not be prone to better trading performance necessarily.

\section{Market impact}

\subsection{Agent learning rate}

\begin{figure}[!htbp]
\begin{centering}
\includegraphics[scale=0.1]{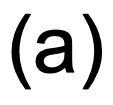}
\includegraphics[scale=0.53]{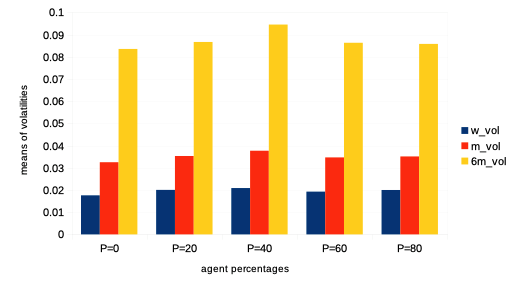}\\
\includegraphics[scale=0.1]{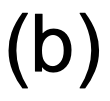}
\includegraphics[scale=0.53]{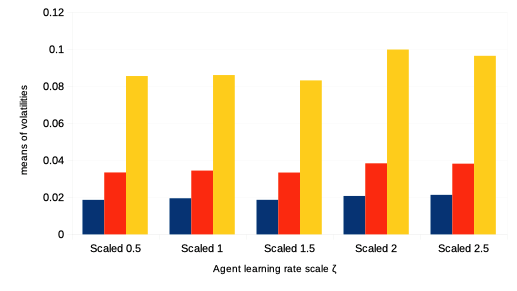}
\caption{\label{N1} (a) Means of volatilities (defined as standard deviations of price normalised to price itself $\sigma/P(t)$) computed over lags of one week (blue), one month (red), and six months (yellow) intervals, for a percentage $p$ of agents corresponding to $p=0\%, 20\%, 40\%, 60\%, 80\%$ of the total agent population with a learning rate scaled by a factor $\zeta=2$ (the remainder $100-p$ being agents initialised with a learning rate $\beta \sim \mathcal{U} (0.05, 0.20)$). (b) Means of volatilities (defined as standard deviations of price normalised to price itself $\sigma/P(t)$) computed over lags of one week (blue), one month (red), and six months (yellow) intervals, for simulations where the entire agent population has a learning rate scaled with factor $\zeta=0.5, 1.0, 1.5, 2.0, 2.5$. The simulations are generated with parameters $I=500$, $J=1$, $T=2875$, $S=20$.}
\end{centering}
\end{figure}

\begin{figure}[!htbp]
\begin{centering}
\includegraphics[scale=0.1]{a.png}
\includegraphics[scale=0.53]{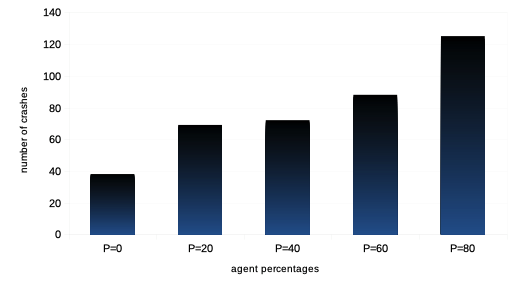}\\
\includegraphics[scale=0.1]{b.png}
\includegraphics[scale=0.53]{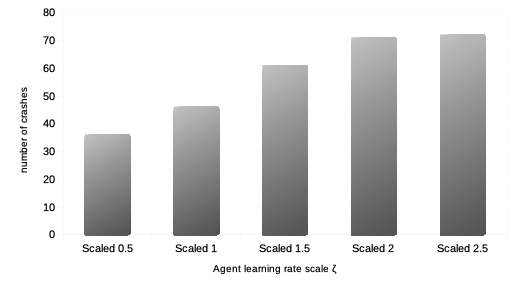}
\caption{\label{N2} (a) Number of market crashes (defined as a drop of more than $20\%$ in market price), for a percentage $p$ of agents corresponding to $p=0\%, 20\%, 40\%, 60\%, 80\%$ of the total agent population with a learning rate scaled by a factor $\zeta=2$ (the remainder $100-p$ being agents initialised with a learning rate $\beta \sim \mathcal{U} (0.05, 0.20)$). (b) Number of market crashes for simulations where the entire agent population has a learning rate scaled with factor $\zeta=0.5, 1.0, 1.5, 2.0, 2.5$. The simulations are generated with parameters $I=500$, $J=1$, $T=2875$, $S=20$.}
\end{centering}
\end{figure}

\begin{figure}[!htbp]
\begin{centering}
\includegraphics[scale=0.1]{a.png}
\includegraphics[scale=0.53]{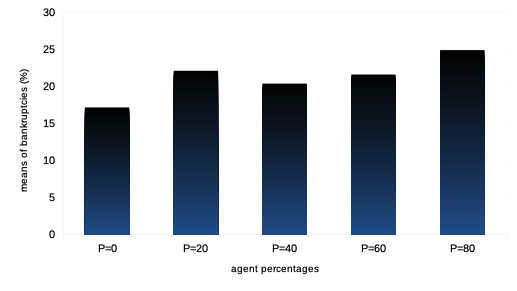}\\
\includegraphics[scale=0.1]{b.png}
\includegraphics[scale=0.53]{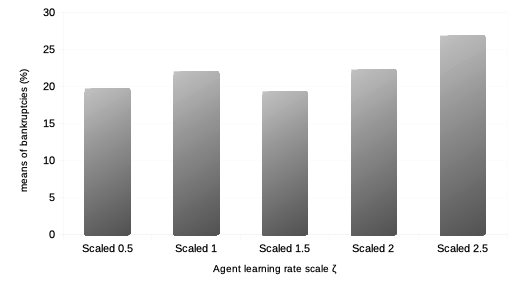}
\caption{\label{N3} (a) Means of all percentage of bankrupt agents at each time step $t$, for simulations with a percentage $p$ of agents corresponding to $p=0\%, 20\%, 40\%, 60\%, 80\%$ of the total agent population with a learning rate scaled by a factor $\zeta=2$ (the remainder $100-p$ being agents initialised with a learning rate $\beta \sim \mathcal{U} (0.05, 0.20)$). (b) Means of all percentage of bankrupt agents at each time step $t$, for simulations where the entire agent population has a learning rate scaled with factor $\zeta=0.5, 1.0, 1.5, 2.0, 2.5$. The simulations are generated with parameters $I=500$, $J=1$, $T=2875$, $S=20$.}
\end{centering}
\end{figure}

We then want to see the impact of the reinforcement learning rate on agent performance and overall market dynamics. Such study is related to the market impact over the years of ever higher frequency and lower latency strategies of trading. Recall each agent is initialised with a learning rate modelled by a parameter $\beta \sim \mathcal{U} (0.05, 0.20)$ for both reinforcement algorithms $\mathcal{F}^{i}$ and $\mathcal{T}^{i}$. In order to do this, we first vary the percentage $p=0\%, 20\%, 40\%, 60\%, 80\%$ of agents with such a learning rate multiplied by a scalar $\zeta=2$, so that it is statistically twice larger than that of the other agents.Then we study the impact of the learning rate when it is scaled by an increasing value of $\zeta=0.5, 1.0, 1.5, 2.0, 2.5$ for the entire agent population. For both such variations in quantity $p$ and quality $\zeta$ of agent learning rates, we can observe the following: 
\begin{itemize}
\item[--] We see on Fig. \ref{N1} rather stable price volatilities at different time-scales.
\item[--] We see on Fig. \ref{N2} a strong increase in the number of market crashes.
\item[--] We see on Fig. \ref{N3} mildly increasing percentages of agent bankruptcies.
\end{itemize}

It is thus interesting to note that both types of variations in agent learning rates in our model do not affect general market volatility, except in tail events with statistically much greater numbers of market crashes. Means of agent bankruptcies are mildly affected.

\subsection{Impact of best agent herding} 

\begin{figure}[!htbp]
\begin{centering}
\includegraphics[scale=0.1]{a.png}
\includegraphics[scale=0.53]{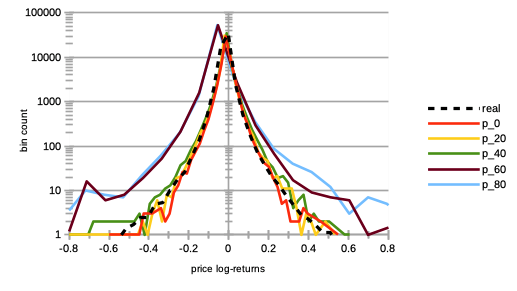}\\
\includegraphics[scale=0.1]{b.png}
\includegraphics[scale=0.53]{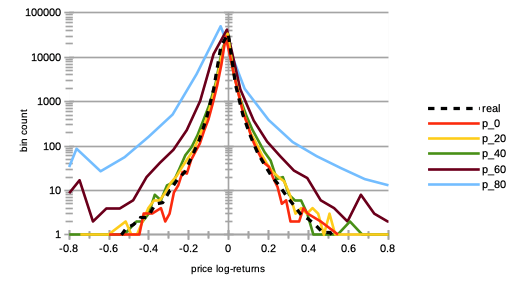}\\
\includegraphics[scale=0.1]{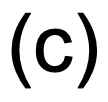}
\includegraphics[scale=0.53]{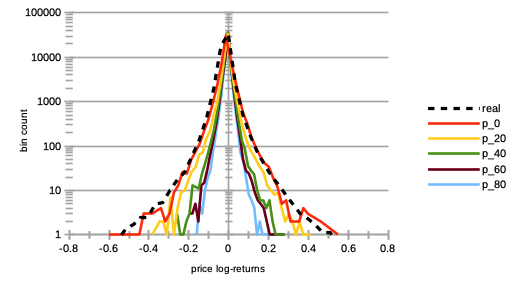}
\caption{\label{L1} Distribution of logarithmic returns of prices $\log [P(t)/P(t-1)]$ of real (dashed black curve) and simulated (continuous curves) data. The simulations are for a percentage $p$ of agents corresponding to $p=0\%$ (red), $p=20\%$ (yellow), $p=40\%$ (green), $p=60\%$ (brown), and $p=80\%$ (light blue) of the total agent population at time $t$ following the best agent (a), following the worst agent (b), or trading randomly (c), while the remainder $100-p$ engaging in proprietary trading strategies. The simulations are generated with parameters $I=500$, $J=1$, $T=2875$, $S=20$.}
\end{centering}
\end{figure}

\begin{figure}[!htbp]
\begin{centering}
\includegraphics[scale=0.53]{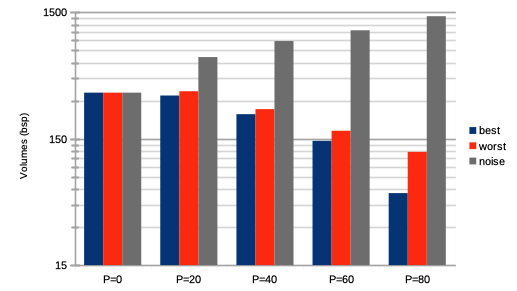}
\caption{\label{L2} Means of all trading volumes, for a percentage $p$ of agents corresponding to $p=0\%, 20\%, 40\%, 60\%, 80\%$ of the total agent population at time $t$ following the best agent (blue), following the worst agent (red), or trading randomly (grey), while the remainder $100-p$ engaging in proprietary trading strategies. The simulations are generated with parameters $I=500$, $J=1$, $T=2875$, $S=20$.}
\end{centering}
\end{figure}

\begin{figure}[!htbp]
\begin{centering}
\includegraphics[scale=0.53]{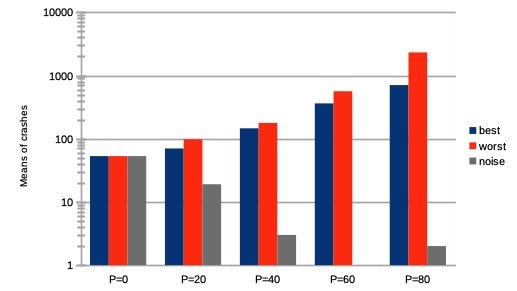}
\caption{\label{L3} Means of market crashes (defined as a drop of more than $20\%$ in market price), for a percentage $p$ of agents corresponding to $p=0\%, 20\%, 40\%, 60\%, 80\%$ of the total agent population at time $t$ following the best agent (blue), following the worst agent (red), or trading randomly (grey), while the remainder $100-p$ engaging in proprietary trading strategies. The simulations are generated with parameters $I=500$, $J=1$, $T=2875$, $S=20$.}
\end{centering}
\end{figure}

\begin{figure}[!htbp]
\begin{centering}
\includegraphics[scale=0.53]{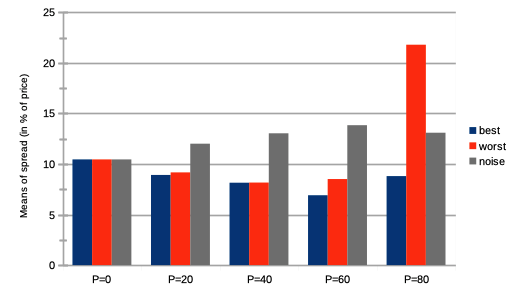}
\caption{\label{L4} Means of bid-ask spread in percentages of the price, for a percentage $p$ of agents corresponding to $p=0\%, 20\%, 40\%, 60\%, 80\%$ of the total agent population at time $t$ following the best agent (blue), following the worst agent (red), or trading randomly (grey), while the remainder $100-p$ engaging in proprietary trading strategies. The simulations are generated with parameters $I=500$, $J=1$, $T=2875$, $S=20$.}
\end{centering}
\end{figure}

The importance of reflexivity in the agent proprietary price estimation and hence trading yield an open question, namely as to what or who is the source of the reflexivity trend endogenous to the market. A first natural and logical answer to this would be renown investors, traders, analysts, etc. that often publish investment recommendations or reviews. We thus can study the impact of agent reflexivity or herding on the market as a whole, as we introduce increasing percentages $p$ of agents sending (when possible) the same transaction order to the order book at time $t+1$ that was sent by the agent with \textit{best} trading performance or track record at time $t$. For the sake of simplicity, we consider this agent with best trading performance as the one with largest net asset value at time $t$. Therefore, the rest of the herding agents may follow and emulate different agents over time (just as in real markets). In particular, for larger percentages of such best herding agents, we can mention the following: 
\begin{itemize}
\item[--] We see on Fig. \ref{L1} a strong increase in price volatilities, especially for higher percentages. Notice that this trend is almost imperceptible for $p<50$.
\item[--] We see on Fig. \ref{L2} extremely decreasing trading volumes.
\item[--] We see on Fig. \ref{L3} extremely increasing numbers of market crashes.
\item[--] We see on Fig. \ref{L4} steadily decreasing market bid-ask spreads, until $p>60\%$, after which they slightly increase again.
\item[--] Remarkably, the rates of agent bankruptcy remain stable regardless of these varying percentages, with average means of $22.76 \pm 3.25\%$ for all values of $p$. 
\end{itemize}

This may be counter-intuitive, but following a renown investor according to this model is extremely averse to market stability.

\subsection{Impact of worst agent herding}

We then want to study the impact of agent reflexivity or herding on the whole market, as we introduce an increasing percentage of agents sending (when possible) the same transaction order to the order book at time $t+1$ that was sent by the agent with \textit{worst} trading performance at time $t$. The interest of this is to study how asymmetric market dynamics become to best agent herding. Here we consider this agent with worst trading performance as the one with lowest, non-bankrupt, net asset value at time $t$. Therefore, the rest of the herding agents may follow and emulate different agents over time. In particular, for larger percentages of such worst herding agents, we can mention the following: 
\begin{itemize}
\item[--] We see on Fig. \ref{L1} a very strong increase in price volatilities, especially for higher percentages. Notice that this trend is almost imperceptible for $p<50$.
\item[--] We see on Fig. \ref{L2} a very strong decrease in trading volumes.
\item[--] We see on Fig. \ref{L3} an extreme increase in market crashes, especially for higher percentages.
\item[--] We see on Fig. \ref{L4} steadily decreasing bid-ask spreads, except for a strong surge for higher percentages $p>80\%$.
\item[--] As one could expect, the rates of agent bankruptcy greatly increase with these varying percentages, staying above $70\%$ of agent bankruptcy for $p>20\%$.
\end{itemize}

We conclude that market instability explodes with increasing proportions of such somewhat unrealistic agents, since no real investor will try and emulate the worst agent. Nevertheless, this shows and validate the previous observations with increasing percentages of agents following the best investor at time $t$.

\subsection{Impact of noise traders}

\begin{figure}[!htbp]
\begin{centering}
\includegraphics[scale=0.53]{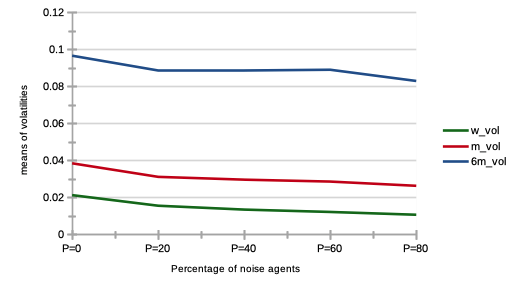}
\caption{\label{K5} Means of volatilities (defined as standard deviations of price normalised to price itself $\sigma/P(t)$) computed over lags of one week (green), one month (red), and six months (blue) intervals, for a percentage $p$ of agents corresponding to $p=0\%, 20\%, 40\%, 60\%, 80\%$ of the total agent population trading randomly (the remainder $100-p$ engaging in proprietary trading strategies). The simulations are generated with parameters $I=500$, $J=1$, $T=2875$, $S=20$.}
\end{centering}
\end{figure}

\begin{figure}[!htbp]
\begin{centering}
\includegraphics[scale=0.53]{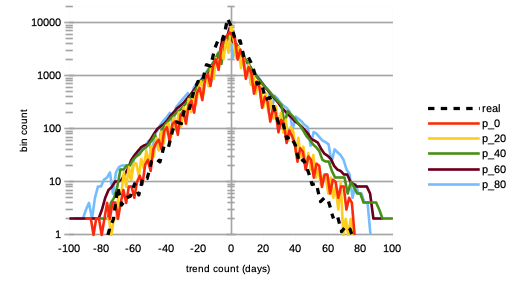}
\caption{\label{K10} Distribution of the number of consecutive days of rising prices (positive values) and dropping prices (negative values). This is for both real (dashed black curve) and simulated (continuous curves) data, the latter being for a percentage $p$ of agents corresponding to $p=0\%$ (red), $p=20\%$ (yellow), $p=40\%$ (green), $p=60\%$ (brown), and $p=80\%$ (light blue) of the total agent population trading randomly (the remainder $100-p$ engaging in proprietary trading strategies). The simulations are generated with parameters $I=500$, $J=1$, $T=2875$, $S=20$.}
\end{centering}
\end{figure}

We then want to study the whole impact of agent learning on the market as we introduce an increasing percentage of ``noise traders", i. e. agents trading randomly~\cite{Schmitt2012}. \textit{A priori}, an ever increasing number of noise agents should bring a certain financial stability to the whole market, by providing more liquidity and higher trading volumes in both bid and offer. In particular, for larger percentages of such ``noise" agents, we can observe the following: 
\begin{itemize}
\item[--] We see on Fig. \ref{L1} a very strong decrease in price volatilities, as seen from price returns distributions, and as we see on Fig. \ref{K5}, a general decrease in volatilities at all time scales.
\item[--] We see on Fig. \ref{L2} a very strong increase in trading volumes.
\item[--] We see on Fig. \ref{L3} a very sharp decrease in market crashes, which virtually almost vanish for $p>50\%$.
\item[--] We see on Fig. \ref{L4} a slow and steady increase in bid-ask spreads.
\item[--] We see on Fig. \ref{K10} a steady increase in length of both bull and bear market regimes, especially the former.
\item[--] Bankruptcy rates steadily decrease with such higher proportion of noise traders from means of $23.22\%$ for $p=0\%$, to $18.84\%$ for $p=80\%$. This is remarkable, as one could have posited that agent survival rates would decrease because of such random trading. 
\end{itemize}

We conclude that counter-intuitively, larger numbers of agents trading randomly is beneficial to market stability and performance.

\section{Conclusion}

Following calibration performances shown in a previous work~\cite{Lussange2019}, we have used a multi-agent reinforcement learning system to model stock market price microstructure. The advantage of such a framework is that it allows to gauge and quantify agent learning, which is at the source of the price formation process, itself at the foundation of all market activity. We first studied agent learning, and then its mesoscale impact on market stability and agent performance. 

According to our results on policy learning, we posit that there are more trading strategies that yield successful portfolio performance, than unsuccessful. 

We also found that best performing agents have a propensity for being fundamentalists rather than chartists in their approach to asset price valuation, and to be less stringent in their choices of transaction orders (i. e. willing to transact orders with larger bids or lower asks). 

Next, we studied the impact on the market of agent learning rates, and found that market volatilities at all time-scales did not vary much (with more agents with larger learning rates, or when all agent collectively have larger learning rates), except for tail events, with average numbers of crashes greatly increasing. We also found that agent bankruptcy rates were not much impacted by such variations in reinforcement learning rates. 

Then we studied the effect of herding or reflexivity, when increasing percentages of agents follow and emulate the investments of the best (and worst) performing agent at each simulation time step. As expected, we found that both such behaviours greatly increase market instability. Yet remarkably, bankruptcy rates of simulations with a best agent herding set up remain quite stable, regardless of the percentages of such agents, and regardless of the yet strongly increasing market volatilities and numbers of crashes. 

Finally, we sought to explore the impact of agent trading information on the price formation process, with larger proportions of ``noise traders" (i. e. agents trading randomly), and found a much greater market stability with increasing percentages of such agents, with a number of crashes virtually vanishing. We also found such markets to be more prone to display bull regimes, and that agents bankruptcy rates slightly diminished. 

We trust such predictions under our model assumptions would be of interest not only to academia, but to industry practitioners and market regulators alike. A natural extension of our model would be to endow agents with a short selling ability (in order to account for specific microstructure effects in times of bubbles for instance). One could also add to each agent to capacity to perform proper portfolio diversification in the model's multivariate framework. Finally, we graciously acknowledge this work was supported by the RFFI grant nr. 16-51-150007 and CNRS PRC nr. 151199.

%\clearpage
\medskip

\bibliography{Article}
 
\end{document}